\documentclass{iau_FM}
\usepackage{graphicx}

\usepackage{comment}

\newcommand{\lsim}{\lower.7ex\hbox{$\;\stackrel{\textstyle<}{\sim}\;$}}

\newcommand {\kms} {\,{\rm km\,s}^{-1}}

\newcommand {\mnras}{\,{\rm MNRAS}}
\newcommand {\aap}{\,A\&A}
\newcommand {\aapr}{\,A\&ARv}
\newcommand {\apjl}{\,ApJL}
\newcommand {\apj}{\,ApJ}
\newcommand {\araa}{\,ARA\&A}
\newcommand {\aj}{\,AJ}
\newcommand {\apjs}{\,ApJS}
\newcommand {\assl}{\,ASSL}
\newcommand {\an}{\,AN}

\title[Angular Momentum Accretion onto Disc Galaxies] 
      {Angular Momentum Accretion onto Disc Galaxies}

\author[Filippo Fraternali \& Gabriele Pezzulli]   
{Filippo Fraternali$^1$
 \and Gabriele Pezzulli$^2$}

\affiliation{$^1$Kapteyn Astronomical Institute, University of Groningen, \\ P.O. Box 800, 9700AV Groningen, The Netherlands,
\\ email: {\tt fraternali@astro.rug.nl} \\[\affilskip]
$^2$Department of Physics, ETH Zurich, \\ Wolfgang-Pauli-Strasse 27, 8093 Zurich, Switzerland
\\email: {\tt gabriele.pezzulli@phys.ethz.ch}}

\pubyear{2018}
\jname{Astronomy in Focus, Volume 1} 
\editors{Teresa Lago \& Danail Obreschkow, ed.}
\doi{10.5281/zenodo.1481543}

\begin{document}

\maketitle

\begin{abstract}
Throughout the Hubble time, gas makes its way from the intergalactic medium into galaxies fuelling their star formation and promoting their growth. One of the key properties of the accreting gas is its angular momentum, which has profound implications for the evolution of, in particular, disc galaxies. Here, we discuss how to infer the angular momentum of the accreting gas using observations of present-day galaxy discs. We first summarize evidence for \emph{ongoing} inside-out growth of star forming discs. We then focus on the chemistry of the discs and show how the observed metallicity gradients can be explained if gas accretes onto a disc rotating with a velocity $20-30\%$ lower than the local circular speed. We also show that these gradients are incompatible with accretion occurring at the edge of the discs and flowing radially inward. Finally, we investigate gas accretion from a hot corona with a cosmological angular momentum distribution and describe how simple models of rotating coronae guarantee the inside-out growth of disc galaxies. 
  
\end{abstract}

\firstsection 

\section{Introduction}\label{sec:intro}

At variance with massive quiescent ellipticals, which assembled most of their mass a long time ago and experienced, at some points in the past, an abrupt decline of their star formation rate, the majority of presently star forming galaxies have been undergoing, for most of the cosmic time, a rather constant or gently declining star formation history (e.g.\ \cite{Pacifici+16}).
This could in principle be explained either by a gradual consumption of a very large initial amount of cold gas, or by continuous accretion of new gas from the intergalactic medium.
Both theory and observations strongly argue in favour of the second option, as i) gradual accretion is expected from the cosmological theory of structure formation (e.g.\ \cite{vdBosch+2014}); ii) star forming galaxies have relatively short depletions times (\cite{Saintonge+2011}) and iii) preventing a huge initial reservoir of gas from very rapid exhaustion requires an implausibly low star formation efficiency, in stark contrast with observations (\cite{Kennicutt&Evans2012}; \cite{Fraternali&Tomassetti2012}).

Observing gas accretion into galaxies directly has proven challenging (\cite{Sancisi+2008}; \cite{Rubin+2012}).
In a galaxy like the present-day Milky Way gas accretion does not seem to take place in the form of cold gas clouds at high column densities, like the classical high-velocity clouds (\cite{Wakker&vanWoerden1997}) as their estimated accretion rate is too low (\cite{Putman+2012}) and their origin may be, at least partially, from a galactic fountain rather than a genuine accretion (\cite{Fraternali+2015}, \cite{Fox+2016}).
Lower column densities have been probed in absorption but the accretion rates are more uncertain (\cite{Lehner&Howk2011}; \cite{Tumlinson+2017}).
A possibility is that gas accretion takes place from the cooling of the hot gas, the galactic coronae that surround the Milky Way and similar galaxies (\cite{Miller&Bregman2015}).
The cooling can be stimulated by the mixing with the disc gas through fountain condensation (\cite{Armillotta+2016}; \cite{Fraternali2017}).
Alternatively, it may come from cold cosmological filaments directly reaching the discs (e.g.\ \cite{Keres+2009}).
To distinguish between these scenarios it is crucial to estimate the properties of the accreting gas.

A powerful tool to investigate the properties of the accreting gas is to infer them indirectly (backward approach) from observations of galaxy discs today.
A simple example of this backward approach consists in the estimate of the accretion rates from the star formation histories (e.g.\ \cite{Fraternali&Tomassetti2012}).
More elaborate estimates allow us to derive the angular momentum of the gas, the location where the accretion should take place and the properties of the medium from which the accretion originates.
These topics are the focus of this proceeding.

\section{Accretion of angular momentum on star forming galaxies}\label{sec:AMaccr}

A very important observational fact about the evolution of currently star forming spiral galaxies, is that they have been increasing in size while increasing in mass (\emph{inside-out growth}, e.g.\ \cite{Larson1976}; \cite{Dale+2016}). This is most likely due to the fact that the gas that has been accreted most recently is more rich in angular momentum with the respect to the one which was accreted at earlier epochs, a very well established prediction of the cosmological theory of tidal torques (\cite{Peebles1969}).

\begin{figure}[bt]
  \begin{center}
    \includegraphics[width=0.54\textwidth]{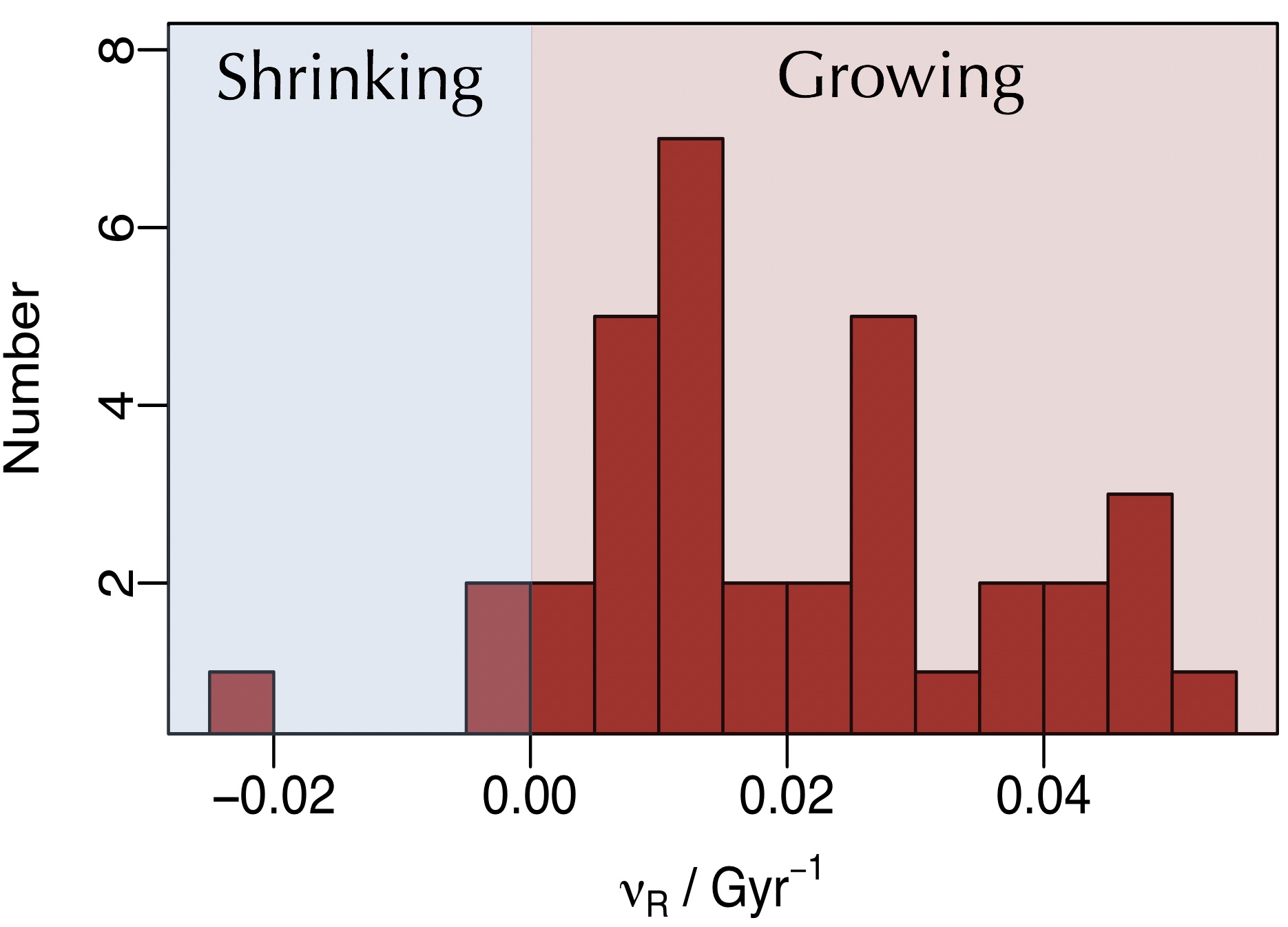} 
    \caption{Distribution of the observed specific radial growth rates ($\nu_R$) of a sample of nearby galaxies (\cite{Pezzulli+2015}). Note how the vast majority are growing inside out still at $z=0$.}
    \label{fig:growing}
  \end{center}
\end{figure}

Crucially, observations indicate the radial growth -- and therefore angular momentum accretion -- is also a gentle process, which has been proceeding at a regular rate throughout galaxy evolution and is still ongoing today, as shown by studies of spatially resolved stellar populations (e.g.\ \cite{Williams+2009}; \cite{Gogarten+2010}) or recent star formation (e.g.\ \cite{Munoz-Mateos+2007}). \cite[Pezzulli et al.\ (2015)]{Pezzulli+2015} have proposed a quantitative analysis of the phenomenon. They have shown that, \emph{relative to} the well known exponential profile of the stellar mass surface density of spiral galaxies, the radial profile of the current \emph{star formation rate surface density} shows a mild depletion in the inner regions and a slight enhancement in the outer ones, which agree both qualitatively and quantitatively with ongoing radial growth of stellar discs at a low but measurable rate.
Figure \ref{fig:growing} shows the distribution of the measured specific radial growth rate $\nu_R\equiv {\dot R_{\star}}/{R_{\star}}$ of the stellar scale-length $R_\star$ of the sample of nearby spiral galaxies from that study.
The vast majority of objects are currently growing, at a rate about equal to one third of their \emph{specific star formation rate} (sSFR, or $\nu_M\equiv {\dot M_{\star}}/{M_{\star}} \simeq 0.1 \; \textrm{Gyr}^{-1}$ at $z=0$, e.g.\ \cite{Speagle+2014}).
Furthermore, the results were shown to agree quantitatively with expectations for gradual angular momentum assembly of galaxies evolving \emph{along} the specific angular momentum versus stellar mass (Fall) relation (\cite{Fall&Romanowsky2013}).

\section{Disentangling models of accretion}\label{sec:chemistry}

We have seen that most spiral galaxies must have been (and probably are) gradually accreting angular momentum rich gas from the surrounding medium. The compelling question arises of what is the exact physical mechanism by which this happens.
Two competing scenarios exist: \emph{cold mode} accretion and \emph{hot mode} accretion (e.g.\ \cite{Birnboim&Dekel2003}; \cite{Binney2004}) and different modes can dominate at different masses and redshift.
In the former case, cold and angular momentum rich gas from intergalactic filaments joins the disc at large radii and then somehow drifts inwards to sustain star formation with the observed radial profile throughout the disc.
We can call this \emph{purely radial} accretion.
In the second scenario, instead, the gas accreting onto the halo does not join the main body of the galaxy \emph{directly}, but it is rather stored (together with its angular momentum) into a hot CGM (\emph{corona}) and then only gradually condenses on to the disc as a gentle `rain', which may be modelled, at least at first order, as mostly vertical accretion (perpendicular to the galaxy disc).

\begin{figure}[bt]
\begin{center}
  \includegraphics[width=0.69\textwidth]{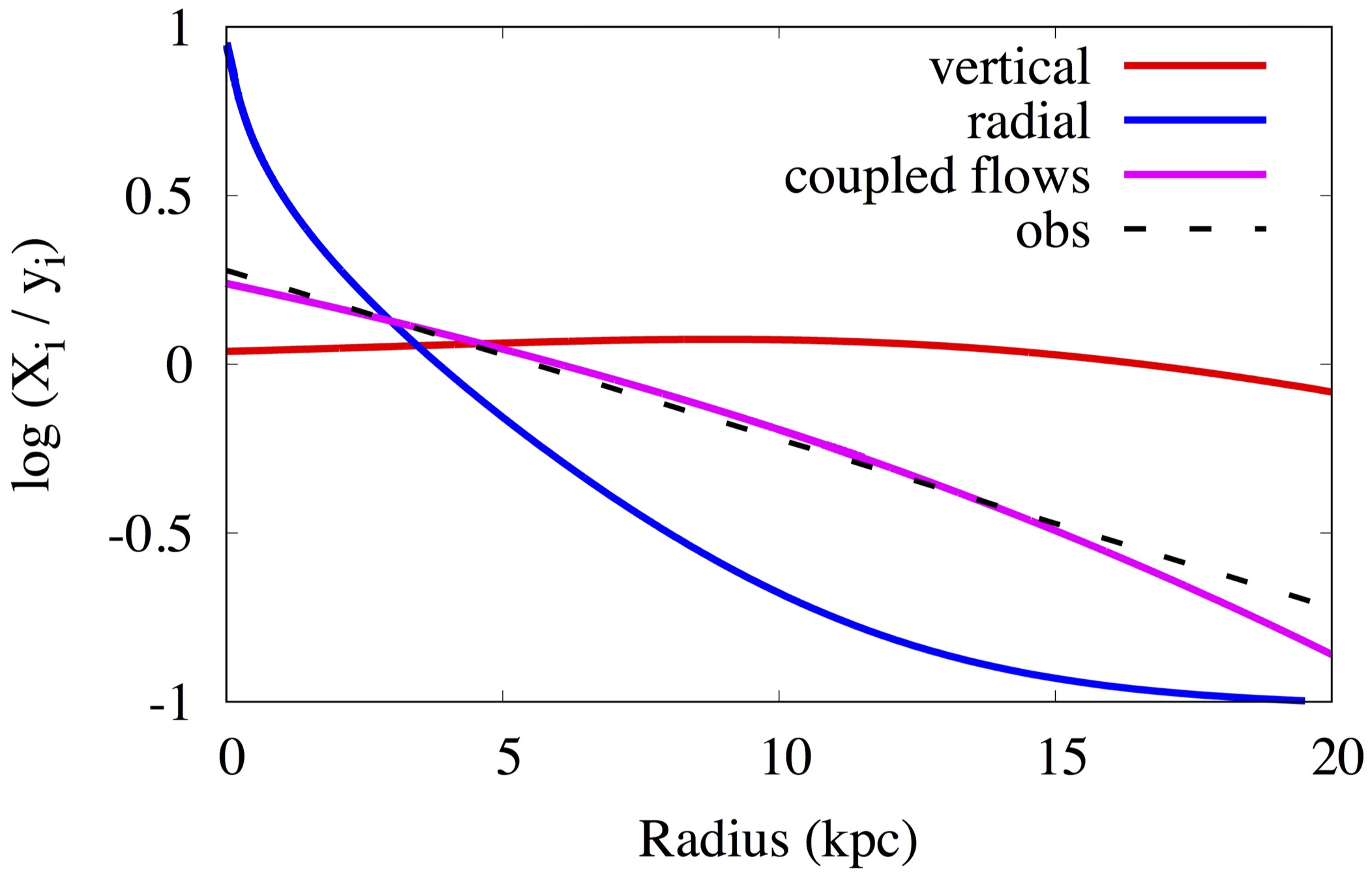} 
 \caption{Abundance profiles of gas phase $\alpha$-elements, as a function of the galactocentric radius, for the Milky Way at $z=0$ predicted by disc evolution models in which the gas accretes purely vertically, purely radially or through mixed (coupled) flows. The observed profile is shown by the dashed curve. The models are calculated as described in \cite{Pezzulli&Fraternali2016b}.}
   \label{fig:gradients}
\end{center}
\end{figure}

With appropriate choices of the parameters, both scenarios can give rise to the same \emph{structural} evolution of the disc, as constrained by observations of the star and gas content of galaxies as a function of galactocentric radius and time (\cite{Pezzulli&Fraternali2016}).
The two models however differ enormously (and can thus be distinguished) in terms of \emph{chemical} evolution.
This is because vertical accretion of relatively metal-poor gas has a \emph{metal-dilution} effect, which goes in the direction of counter-acting metal enrichment by local star formation, whereas radial accretion implies that, before arriving at the position where it is finally locked in to stars, each gas element will have already traversed other regions of the galaxy, where it will have been chemically enriched by the stars being formed there.
We emphasize that i) this observational test is better performed on \emph{gas-phase} abundances of $\alpha$ elements (as this choice minimizes uncertainties due to stellar radial migration and time delays in chemical enrichment) and ii) this kind of comparison between models is only meaningful \emph{at fixed structural evolution}, as, otherwise, differences in other leading order effects (gas fraction, star formation efficiency and so on) dominate over those due to different geometry of accretion.
With these specifications clarified, the discriminating power of the method is remarkable.
This is illustrated in Figure \ref{fig:gradients}, where the predictions are shown, for the abundance gradient of $\alpha$-elements in the ISM of the Milky Way, for models with purely vertical, purely radial and mixed accretion. Details can be found in e.g.\ \cite{Pitts&Tayler1989}, \cite{Schoenrich&Binney2009} and \cite[Pezzulli \& Fraternali (2016a)]{Pezzulli&Fraternali2016}.
The latter work proposes an analytic and general approach to the problem, which can be readily applied to any galaxy or structural evolution model.

The clear result is that a combination of vertical and radial accretion is required to match the observed gradient (e.g.\ \cite{Genovali+15}, marked here as a dashed line).
This is actually not surprising, when angular momentum conservation is taken into account.
A purely radial accretion, in fact, requires the angular momentum of the accreting gas to be transferred to some other not very well identified phase. On the other hand, purely vertical accretion is only possible if the material is accreted, at any radius, with exactly the angular momentum needed for local centrifugal balance, as any discrepancy would force the condensed gas to move radially within the disc after accretion (see also \cite{Mayor&Vigroux1981}; \cite{Bilitewski&Schoenrich2012}). We now discuss whether a coherent physical picture can naturally give account of the findings discussed so far.

\section{A consistent picture}

The model which better reproduces the chemical evolution of the Milky Way (\S\ref{sec:chemistry}) requires gas accretion occurring with a specific angular momentum that is $75\pm 5 \%$ of that of the disc at each radius. This is very naturally expected for a hot mode accretion scenario (i.e.\ from the hot corona).
Virtually every hydrodynamically consistent model of the hot CGM requires in fact the hot gas to be \emph{not} in local centrifugal equilibrium, as the high temperatures will generally imply a significant contribution of pressure support (in addition to rotation) against gravity. Note that the same model also predicts the presence of moderate radial flows (a few $\kms$ or less) within the disc: crucially, however, this radial flow is not due to equatorial accretion of cold flows, but it is rather the natural consequence of the rotation lag of the accreting (hot) gas and angular momentum conservation.

Two main questions arise to further test whether the model is viable in a cosmological context. First, if the disc accretes material with a local deficit of angular momentum, can the accretion still \emph{globally} provide enough angular momentum, as required to sustain the global radial growth of the disc (\S\ref{sec:AMaccr})? Second, is the implied rotation of the corona consistent with cosmological expectation from tidal torque theory?

The answer to these questions requires building self-consistent models of the hydrodynamical equilibrium of a hot rotating corona in a galaxy scale gravitational potential and with a given angular momentum distribution. \cite{Pezzulli+17} described the solution to this problem and showed that a corona with a cosmologically motivated angular momentum distribution can naturally develop, \emph{in the proximity of the disc}, rotation velocities close to the value required to match the chemical constraints.
The model also predicts that the rotation velocity of the hot gas should drop significantly when approaching the virial radius.The first prediction is in excellent agreement with the recent observations by \cite{Hodges-Kluck+2016}; the second will require next-generation X-ray observations to be confirmed or discarded.

\cite{Pezzulli+17} also found that the specific angular momentum of the inner corona increases rather steeply with radius and becomes larger than the \emph{average} angular momentum of the disc at a radius $R_{\rm crit}$ slightly larger than the disc scale-length, but well within the range of direct contact between the galaxy and the hot halo. This is is sufficient to make the corona a plausible source of angular momentum growth, provided that the accretion of coronal gas is particularly efficient at relatively large radii, as predicted for instance by models of fountain-driven condensation (e.g.\ \cite{Marasco+2012}; \cite{Fraternali+2013}), and/or that the accretion is inhibited or counter-acted in the very central regions by star formation or AGN feedback (as suggested for instance for the Milky Way by the discovery of the Fermi bubbles; \cite{Su+2010}).


\end{document}